\documentclass[a4paper]{jpconf}

\def\AEF{Faraggi A E}

\usepackage{amsmath,amssymb,lmodern}
\usepackage{graphicx}
\usepackage{amsfonts}
\usepackage{bm}

\def\NPB#1#2#3{#2 {\it Nucl.\ Phys.}\/ {\bf B#1} #3}
\def\PLB#1#2#3{#2 {\it Phys.\ Lett.}\/ {\bf B#1} #3}
\def\PLA#1#2#3{#2 {\it Phys.\ Lett.}\/ {\bf A#1} #3}
\def\PRD#1#2#3{#2 {\it Phys.\ Rev.}\/ {\bf D#1}  #3}
\def\PRL#1#2#3{#2 {\it Phys.\ Rev.\ Lett.}\/ {\bf #1} #3}
\def\PRT#1#2#3{#2 {\it Phys.\ Rep.}\/ {\bf#1} #3}

\def\IJMP#1#2#3{#2 {\it Int.\ J.\ Mod.\ Phys.}\/ {\bf A#1} #3}

\def\EJP#1#2#3{#2 {\it Eur.\ Phys.\ Jour.}\/ {\bf C#1} #3}

\def\JCA#1#2#3{#2 {\it JCAP}\/ {\bf #1} #3}

\newcommand{\beq}{\begin{equation}}
\newcommand{\eeq}{\end{equation}}
\newcommand{\beqa}{\begin{eqnarray}}
\newcommand{\beqn}{\begin{eqnarray}}
\newcommand{\eeqn}{\end{eqnarray}}
\newcommand{\eeqa}{\end{eqnarray}}

\usepackage{epsf}
\usepackage{graphicx}
\begin{document}
\title{Hamilton--Jacobi meet M\"obius}

\author{$\underline{\hbox{Alon E. Faraggi}}^1$ and Marco Matone$^2$}

\address{$^1$ Department of Mathematical Sciences, University of Liverpool, 
Liverpool L69 7ZL, UK}

\address{$^2$ Dipartimento di Fisica, Universit`a di Padova, 
Via Marzolo 8, I--35131 Padova, Italy}

\ead{alon.faraggi@liv.ac.uk; marco.matone@pd.infn.it}

\begin{abstract}

Adaptation of the Hamilton--Jacobi formalism to quantum 
mechanics leads to a cocycle condition, which is 
invariant under $D$--dimensional M\"obius transformations
with Euclidean or Minkowski metrics. In this paper we aim to 
provide a pedagogical presentation of the proof of the M\"obius 
symmetry underlying the cocycle condition. The M\"obius 
symmetry implies energy quantization and undefinability of quantum
trajectories, without assigning any prior interpretation to the 
wave function. As such, the Hamilton--Jacobi formalism,
augmented with the global M\"obius symmetry,
provides an alternative starting point, to the axiomatic 
probability interpretation of the wave function, for the 
formulation of quantum mechanics and the quantum spacetime. 
The M\"obius symmetry can only be implemented consistently
if spatial space is compact, and correspondingly if
there exist a finite ultraviolet length scale. Evidence
for non--trivial space topology may exist in the 
cosmic microwave background radiation.

\end{abstract}

\section{Introduction}

The possibility that the Standard Model of particle physics
provides a viable parameterisation of subatomic data up to the 
Planck scale received substantial support from the 
observation of a Higgs--like particle at the LHC. 
While the properties of this Higgs--like particle will be 
the subject of experimental scrutiny in the decades to come,
it is clear that fundamental understanding of the Standard Model
parameters can only be gained by incorporating gravity into the 
picture. Alas, the synthesis of gravity and quantum mechanics 
remains an enigma. The majority of contemporary efforts entail 
the quantisation of gravity, {\it i.e.} the quantisation
of the metric in Einstein's theory of general relativity. 
The most developed approach in this endeavour is string theory \cite{stheory}.
While many alternative approaches exist, which in principle should
be regarded on equal footing, the key advantage of string
theory is that, while producing a consistent approach 
to quantum gravity, it gives rise to the matter and 
gauge structures that arise in the Standard Model.
Several examples of quasi--realistic string models
have been constructed over the years, and include 
the free fermionic standard--like models \cite{ffslm}.
This class of models produce solely the states 
of the Minimal Supersymmetric Standard Model 
in the Standard Model charged sector, and produced
a successful prediction of the top quark mass \cite{topmass}.
String theory provides an effective perturbative approach 
to quantum gravity. Important properties of string theory
include the various perturbative \cite{gpr}
and non--perturbative dualities \cite{ht}. 
While string theory, due to its relation 
to observational data, can be regarded as 
an important step on the road to the synthesis of
gravity and quantum mechanics, it does not
provide a satisfactory final answer. 
What we would like to have is a formulation 
of quantum gravity, which follows from a basic 
physical hypothesis, akin to equivalence principle 
underlying general relativity. 

An alternative approach to the quantisation of gravity
may be pursued by seeking the geometrisation of quantum 
mechanics. 
This is the aim of the equivalence postulate (EP) approach to
quantum mechanics 
\cite{fm2, fmut, fmqt, fmtqt, fmpl, fmijmp, bfm, fmmt}. 
The EP formalism may be regarded 
as adaptation of the Hamilton--Jacobi (HJ) formalism to 
quantum mechanics. In the classical HJ theory the solution 
to the mechanical problem is obtained by performing
canonical transformation to a system in which the 
phase space variables are constants of the motion. 
The Hamiltonian in the new system therefore vanishes 
identically and the Hamilton equations of motion
are zero. The transformation from the old set of 
to the new set treat the phase space variables 
as independent variables. The solution to this
problem is given by the Classical Hamilton--Jacobi
Equation (CHJE). The functional relation between the
phase space variables is then extracted from the
solution of the CHJE $S(q)$. We may a pose a similar
question, but rather than performing canonical 
transformations that treat the phase space
variables as independent variables, 
we seek a trivialising coordinate transformation from 
the system $q^a$ to the system $q^b$,
and the induced transformation 
$p^a=\partial_a S^a(q^a)\rightarrow p^b=\partial_b S^b(q^b)$. 
Without loss of generality we further impose that $S^a(q^a)=S^b(q^b)$, 
{\it i.e.} that $S(q)$ transform as a scalar function under the transformation. 
By trivialising we mean that in the new coordinate system the kinetic and potential
energy vanish. However, it is clear that this proposition would not 
make sense in classical mechanics. The reason is that in classical 
mechanics we have the solution $S(q)=constant$ for the state
with vanishing kinetic energy and vanishing potential. 
In this case we have $p\equiv 0$, and this will remain the case 
after the transformation. This does not make sense because 
if we insist that all the physical states should be connectable to 
the free state with vanishing energy and vanishing potential,
consistency dictates that the inverse transformation should exist as well.
This means that we have to remove the solution $S(q)=constant$
from the space of allowed solution. We therefore seek a modification of
the CHJE that will remove the state  $S(q)=constant$ from the
space of allowed solutions. In the stationary case
the CHJE is given by
\beq
{1\over{2m}}\left({{\partial{ S}_0}\over%
{\partial q}}\right)^2~~+~~V(q)~~-~~E~~=~~0. 
\label{hjsc}
\eeq
We consider the modification of the CHJE by adding the function 
$Q(q)$ to the CHJE, whose properties are yet to be determined.
We further require that the all the physical systems
labelled by $W(q)=V(q)-E$ can be connected by coordinate transformations
to the trivial state $W(q)\equiv 0$. The condition that $S(q)\ne constant$ 
and that $S(q) = constant$ can be reached only in the classical limit,
entails that the classical limit coincides with the limit $Q(q)\rightarrow 0$. 

Imposing these conditions consistently leads to a cocyle condition. 
The modified Hamilton--Jacobi equation corresponds to the Quantum Hamilton--Jacobi
Equation (QHJE), and is related to the Schr\"odinger equation. 
The cocycle condition is invariant under $D$--dimensional 
M\"obius transformations. This invariance is the key
property of quantum mechanics in the EP approach, 
and is equivalent to the condition that $S(q)\ne constant$. 
The M\"obius symmetry implies that spatial space is compact 
and the the decompactification limit coincides with the 
classical limit, {\it i.e.} the limit in which $Q(q)\rightarrow 0$.
Furthermore, it implies the existence of a finite length scale in 
the ultraviolet and correspondingly a finite length scale in 
the infrared. Hence, the M\"obius symmetry has far reaching 
implications on the structure of the geometry, 
and correspondingly on the structure of the quantum space time. 
In this paper our aim is to provide a pedagogical presentation 
of the proof of the M\"obius invariance of the cocycle 
condition.

\section{The Hamilton--Jacobi Theory}

Physics is first and foremost an experimental
science, and may be defined as the mathematical modelling
of experimental observations. A successful mathematical model
is the one that can account for a wider range of experimental data. 
One may contemplate the possibility that there exist a representation
of the physical laws, which is devoid of any experimental input, 
and follows from rigorous mathematical reasoning. The relation 
between the circumference of a circle to its diameter may attest 
to this possibility. 
Given the finite period allocated to an experiment, 
its resolution is similarly limited. 
Infinite resolution requires infinite time and
the proposition that physics may have a completely
mathematically rigorous representation, without resource 
to experimental input, is theological.

The methodology of the mathematical modelling of experimental
observations is constructed as follows. Start with some
initial conditions of a set of physical observables. 
Build a mathematical model for how this set of variables
may evolve in the course of the physical process. 
Measure the set of variables in some experimental 
apparatus. Confront the observations by using the 
experimental apparatus with the prediction of the 
mathematical model. The key to the experimental 
methodology is that, given a proper set of 
instructions for the preparation of the 
initial conditions and the experimental apparatus, 
the outcome of the experimental measurements will
be identical.

The mathematical modelling therefore rests on 
identifying a set of variables that are to 
be measured in the experiment. In the Newtonian
system these may be the position and velocities 
of some objects. Say, a ball is dropped at some 
height $h$, with zero initial velocity. Then 
measure its position and velocity as a function of time. 
In this type of experiment 
the position and velocity change with time, 
which makes the experiment difficult. What we
would like to do is to measure
quantities that do not change with time. 
For example, if we can measure the initial 
energy and we know that the energy does 
not change with time, then we can measure
the energy at the end of the experiment 
and we should get the same value. We can 
model the physical process from the 
start of the experiment and how the total 
energy is distributed among the constituents 
of the mathematical model. Ultimately, 
whatever happens between the start and 
the end of the experiment the total energy should be 
the same. What we are after are the constants 
of the motion. 
The Hamilton--Jacobi (HJ) formulation of classical mechanics
is completely equivalent to Newtonian mechanics, and  
provides a general method to extract the 
constants of the motion for any physical system. 

To achieve this fiat in classical mechanics we have 
to make a change of variables from configuration space
to phase space. In configuration space the set 
of variables are the positions and velocities. 
In phase space the set of variables are the 
coordinates and the momenta. What we are 
after is to get rid of the explicit time dependence 
of the variables, so that the constants of the 
motion can be more readily extracted. 

The Hamiltonian in classical mechanics is a
function of the phase space variables. 
In systems in which the Hamiltonian 
does not depend explicitly on time 
the Hamiltonian corresponds to the total energy 
of the physical system and is a constant of the 
motion. The Hamilton equations of motion are 
given by 
\beq
~~~~\dot q~=~{\partial H\over\partial p}~~,~~\dot p=-{\partial H\over\partial q}
\label{heom}
\eeq
The Hamilton--Jacobi procedure then follows by transforming 
the Hamiltonian to a trivial Hamiltonian, with the result that 
the new phase space--variables are constants of the motion, {\it i.e.}
$$H(q,p)~~\longrightarrow~~ { K}({ Q},{P})\equiv0~~~~~~~~\Longrightarrow~~~~~\dot{  Q}={\partial{K}\over
{\partial{P}}}\equiv 0~,~\dot{P}=-{\partial {K}\over{\partial Q}}\equiv 0 ~~~~~~~~~
$$
The solution is the Classical Hamilton--Jacobi Equation
$$H(q,p)~\longrightarrow~ {K}({Q},{P})~=~H(q,p={\partial S\over\partial q})~+~
{{\partial S}\over{\partial t}}~=~0~~\Rightarrow~~{\rm CHJE}, ~~~~~$$
which in the stationary case becomes eq. (\ref{hjsc}). The transformations
from the old to the new phase space variables, 
$(q,p)\rightarrow (Q,P)$ are canonical, which treat the phase space
variables as independent variables. Their functional relation is then 
extracted from the solution of the HJ equation as $p=\partial_q S_0(q)$. 
We may reverse this order. Namely, we seek a trivialising coordinate 
transformation, such that $V(q)=E=0$, but keeping the relation 
$p=\partial_q S_0(q)=$ when performing the transformations. 
The transformation are reversible coordinate transformations. 
Generality, then demands that we should be able to perform the 
reverse of the transformation from the trivial state to the 
non--trivial state. Furthermore, from the trivial state, we should 
be able to transform to any non--trivial state. However, 
as we discussed above this is not consistent in classical 
mechanics. As seen from eq. (\ref{hjsc}), while the first term 
transforms as quadratic differential, the second term, in general, 
does not. Furthermore, the state $W(q)=V(q)-E=0$, corresponding 
to the solution $S_0=constant$ with $p=0$, is a fixed point under the 
transformations. Therefore, consistent implementation of this
trivialising procedure requires the modification of the CHJE. 
As we elaborate in the next section the modification leads 
to a cocycle condition which is invariant under $D$--dimensional 
M\"obius transformations. Our aim is to provide a 
pedagogical presentation of this proof. The M\"obius symmetry
is the fundamental property underlying quantum mechanics 
in our approach. It provides an alternative to the axiomatic 
probability interpretation of the wave function, and by that 
it provides a framework for rigorous formulation of quantum
mechanics. Furthermore, invariance under the M\"obius 
transformations reveals the existence of an inherent 
fundamental length in the formalism \cite{fmpl}. 
Proper implementation of the classical limit then shows that this
length scale may be identified with the Planck length \cite{fmpl, fmijmp}. 
In turn, the existence of a minimal length scale and
the M\"obius symmetry implies the existence of a maximal length scale. 
Furthermore, the limit in which the maximal length scale goes to infinity, 
{\it i.e.} the decompactification limit, correspondingly corresponds to 
the classical limit. The M\"obius symmetry underlying 
the Quantum Hamilton--Jacobi Theory (QHJT), therefore
carries within it an intrinsic ultraviolet regularisation 
length and correspondingly an infrared finite scale. The QHJT,
with its underlying M\"obius symmetry, provides the arena 
for the proper understanding of the quantum spacetime.

\section{The cocycle condition}
We seek
the $v$--transformations 
$$(~q~,~S_0(q)~,~p~=~{{\partial S_0}\over{\partial q}})~~\longrightarrow~~
(~{q^v}~,~{S^v}_0({q^v})~,~{ p^v}~=~
{{\partial{S^v}_0}\over{\partial{q^v}}}~)$$
Such that 
$W(q)~~\longrightarrow~~{W^0}({q^0})~=~0$
exist for all $W(q)$. In the following we impose the conditions
  \begin{equation}
  \label{vmap}
  S^v(q^v)=S(q),
  \end{equation}
and in particular
\begin{equation}
\label{vmap0}
S^0(q^0)=S(q).
\end{equation}

The CSHJE for a particle of mass $m$ and energy $E$ moving in $N$
dimensions under the influence of a static velocity independent
potential $V(q)$ is
\begin{equation}
\frac{1}{2m}\sum_{i=1}^N
\left(\frac{\partial S}{\partial q_i}\right)^2 + 
W(q) = 0,
\label{cshje} 
\end{equation}
where $W\equiv V-E$, and where the $q_i$ are Cartesian
coordinates. Under a change of
coordinates $q\rightarrow q^v$ we have by (\ref{vmap})
\begin{equation}
\frac{\partial S^v(q^v)}{\partial q^v_j} =
\frac{\partial S(q)}{\partial q^v_j} = \sum_i \frac{\partial
S(q)}{\partial q_i}\frac{\partial q_i}{\partial q^v_j},
\label{jac} 
\end{equation}
which we can write as
\begin{equation}
\bm{p}^v=\mathbf{J}^v\bm{p}
\label{pdef} 
\end{equation}
 where
$$J^v_{ij}=\frac{\partial q_i}{\partial q^v_j}$$ is the Jacobian
matrix connecting the coordinate systems $q$ and $q^v$ and
$p_i=\frac{\partial S}{\partial q_i}$. Then
\begin{eqnarray}
\nonumber \sum_j \left(\frac{\partial S^v}{\partial
q^v_j}\right)^2&=&|\bm{p}^v|^2\\
\nonumber &=&\left(\frac{|\bm{p}^v|^2}{|\bm{p}|^2}\right)|\bm{p}|^2\\
\label{p2def} &=& (p^v|p)|\bm{p}|^2,
\end{eqnarray}
where we defined
\begin{eqnarray}
\label{trafact} (p^v|p) &\equiv&
\frac{|\bm{p}^v|^2}{|\bm{p}|^2}\\
\nonumber &=&
\frac{{\bm{p}^v}^\mathsf{T}\bm{p}^v}{\bm{p}^\mathsf{T}\bm{p}}\\
\label{jacsquare} &=& \frac{\bm{p}^\mathsf{T}
{\mathbf{J}^v}^\mathsf{T} \mathbf{J}^v \bm{p}}{\bm{p}^\mathsf{T}
\bm{p}}.
\end{eqnarray}
Note also that by (\ref{pdef}) 
\begin{eqnarray} 
\bm{p}&=&{\mathbf{J}^v}^{-1}\bm{p}^v\\
\nonumber \Rightarrow \qquad
p_i&=&\sum_j({\mathbf{J}^v}^{-1})_{ij}p^v_j.
\label{backwds}
\end{eqnarray}
But simply relabelling (\ref{jac})
\begin{equation}
p_i=\frac{\partial S(q)}{\partial q_i}=\frac{\partial
S^v(q^v)}{\partial q_i}=\sum_j \frac{\partial S^v(q^v)}{\partial
q^v_j}\frac{\partial q^v_j}{\partial q_i}=\sum_j \frac{\partial
q^v_j}{\partial q_i}p^v_j.
\end{equation}
Thus $({\mathbf{J}^v}^{-1})_{ij}=\frac{\partial q^v_j}{\partial
q_i}$ and we have
\begin{equation}
\label{jinv}
({\mathbf{J}^v}^{-1})_{ij}=(J^v_{ij})^{-1}.
\end{equation}
 Then using (\ref{backwds}) in (\ref{trafact}) we have
\begin{equation}
\label{vswitch}\left[(p^v|p)\right]^{-1}=(p|p^v)=
\frac{{\bm{p}^v}^{\mathsf{T}}{{\mathbf{J}^v}^{-1}}^{\mathsf{T}}
{\mathbf{J}^v}^{-1}\bm{p}^v} {{\bm{p}^v}^{\mathsf{T}}\bm{p}^v},
\end{equation}
where by (\ref{jinv})
\begin{eqnarray}
\nonumber [{{\mathbf{J}^v}^{-1}}^{\mathsf{T}}
{\mathbf{J}^v}^{-1}]_{jk}&=&\sum_i{\mathbf{J}^v}^{-1}_{ik}
{\mathbf{J}^v}^{-1}_{ij}\\
\label{jav} &=&\sum_i\left[J^v_{ik}J^v_{ij}\right]^{-1}.
\end{eqnarray}

Having determined that the first term
in (\ref{cshje}) transforms as a quadratic differential under
$v$-maps, we require that for consistency the second term 
transforms similarly
That is:
\begin{equation}
W^v(q^v) = (p^v|p)W(q),
\label{Wtran} 
\end{equation}
and
\begin{equation}
W^0(q^0)=(p^0|p)W(q),
\label{W0} 
\end{equation}
where $W^0\equiv 0$ corresponds to the
free particle at rest, with $V=0$ and $E=0$. 
Substituting $W^0=0$ in the left--hand side of 
eq. (\ref{W0}) yields
\begin{equation}
\label{imp} 0=(p^0|p)W(q).
\end{equation}
Hence, $W^0$ is a fixed point under the $v$--maps and 
cannot be connected to other states. 
Therefore, classical mechanics in not compatible
with the equivalence postulate.

As the CSHJE is not consistent with the equivalence postulate,
we consider the modification
$$W\rightarrow W + {Q},$$ 
The Modified Stationary Hamilton Jacobi Equation takes the form
\begin{equation}
\label{QSHJE} \frac{1}{2m}\sum_i \left(\frac{\partial S}{\partial
q_i}\right)^2 + W(q) + {Q}(q) = 0.
\end{equation}
Covariance under $v$--map imposes the transformation property
\begin{equation}
W^v +
{Q}^v=(p^v|p)\left[W + {Q}\right].
\label{WandQ} 
\end{equation}
In order that the $v$--map may connect $W^0$ to any other
state, $W$ must transform with an inhomogeneous term,
\begin{equation}
W^v = (p^v|p)W + (q;q^v), 
\label{inhom} 
\end{equation}
and by eq, (\ref{WandQ}) 
\begin{equation}
{Q}^v = (p^v|p){Q} - (q;q^v).
\label{inhomQ} 
\end{equation}
We note that for an arbitrary state
\begin{eqnarray}
\nonumber W &=& (p|p^0)W^0 + (q^0;q)\\
\nonumber &=& 0 + (q^0;q)\\
&=& (q^0;q),
\label{Wmake} 
\end{eqnarray}
so the inhomogeneous term for the trivialising map generates all
other $W$ states. Let us consider the properties of
the inhomogeneous term. 
First, we note from (\ref{inhom}) 
that in the case of the identity
transformation $q\rightarrow q$
\begin{eqnarray}
W &=& (p|p)W + (q;q)\nonumber \\
\Rightarrow (q;q)&=&0.
 \label{same0}
 \end{eqnarray}
Second, from eq. (\ref{inhom}) we can write
\begin{equation}
W^b = (p^b|p^a)W^a + (q^a;q^b),
\label{ba}
\end{equation}
\begin{equation}
W^c = (p^c|p^a)W^a +
(q^a;q^c),
\label{ca}
\end{equation} 
and
\begin{equation}
W^c =
(p^c|p^b)W^b + (q^b;q^c),
 \label{cb}
\end{equation}
for three states $a,b,c$. Substituting (\ref{ba}) into (\ref{cb})
and equating this with (\ref{ca}) we obtain
\begin{eqnarray}
(p^c|p^a)W^a + (q^a;q^c) &=&
(p^c|p^b)\left[(p^b|p^a)W^a + (q^a;q^b)\right] +
(q^b;q^c)\nonumber \\
\Rightarrow (q^a;q^c) - (q^b;q^c)&=&
(p^c|p^b)(q^a;q^b),
\label{cocyle1} 
\end{eqnarray}
or
\begin{equation}
(q^a;q^b)=(p^b|p^c)\left[(q^a;q^c) -
(q^b;q^c)\right].
\label{cocycle} 
\end{equation}
Eq. {\ref{cocycle}) is our celebrated 
{\it cocycle condition}. It expresses the essence of the 
quantum mechanics in the equivalence postulate approach.
We further note that if we set
$q^a=q^c$ in (\ref{cocycle}) we get
\begin{equation}
\label{swap}
(q^a;q^b)=-(p^b|p^a)(q^b;q^a).
\end{equation}
\section{Higher dimensional M\"{o}bius group}\label{mobiusgroup}
The form of the cocycle condition has far--reaching implications.
Here we reproduce the arguments introduced in
\cite{bfm} which reveal a symmetry of the inhomogeneous
term $(q^a;q^b)$, under $D$--dimensional M\"obius transformations. 
Our aim is to provide a more pedagogical presentation 
of the proof of these properties.

%
 
We denote by $q=(q_1,\cdots,q_D)$ an arbitrary point in 
$R^D$. A similarity is
a $D$--dimensional transformation that includes 
translations, rotations and dilatations. 
Similarities are naturally extended to the compactified space 
$\hat R^D= R^D\cup\{\infty\}$. A similarity maps $\infty$ to itself. 
%
%
Setting    
\beq
r^2=q_1^2+\cdots+q_D^2, 
\label{unitsphere}
\eeq
the last generator of the M\"obius group is the 
{\it inversion} or reflection in the 
unit sphere $S^{D-1}$. If $q\ne0$ 
\beq
q\longrightarrow q^*={q\over r^2}, 
\label{inversion}
\eeq
otherwise 
\beq
0\longrightarrow\infty,\qquad\infty\longrightarrow0. 
\label{inversion2}
\eeq

Summarising,  
an arbitrary
M\"{o}bius transformation of an $N$-dimensional vector
$q=(q_1,\ldots,q_i,\ldots,q_N)$ is made up of any combination of: 

\begin{itemize}
\item a `translation'  ~~~~~$q\rightarrow q^B= q + B$  ~~~~~\vspace{1.8mm}
($B$ a vector in $R^D$), \\
\item a `dilatation'  ~~~~~~$q\rightarrow q^A= Aq$ ~~~~~~~~~\vspace{1.8mm}
($A$ a real number), \\
\item a `rotation' ~~~~~~~ $q\rightarrow q^R= Rq$  ~~~~~~~~~\vspace{1.8mm}
($R$ an orthogonal $D\times D$matrix), \\
\item an `inversion' 
 ~~~~~$q\rightarrow q^{\ast}= \frac{q}{r^2}$ ~~~~~~~~~~
                          \vspace{1.8mm}($r^2\equiv \sum_i q_i^2$).
\end{itemize}

The M\"obius transformations naturally extend to the compactified space 
$\hat R^D= R^D\cup\{\infty\}$. A similarity maps $\infty$ to itself. 
The M\"obius group $M(\hat{R}^D)$ is defined as the set of 
transformations generated 
by all similarities together with the inversion. 
A general M\"obius 
transformation is the combination of a number of reflections and inversions. 
A M\"obius transformation is conformal with respect to the Euclidean metric.
A theorem due to Liouville states that the conformal group.
$M(\hat{R}^D)$ actually coincide\footnote{for review
of the $D$--dimensional M\"obius group, see {\it e.g.} \cite{vuorinen}.}
for $D>2$.

\subsection{Transformation Factors}

In the subsections bellow we derive the transformations factors
$(p^M\vert p)$ in the case of the M\"obius transformations $q\rightarrow q^M$
discussed in section \ref{mobiusgroup}.

\subsubsection{Translation}
The Jacobian for a translation is
\begin{equation}
J^B_{ij}=\frac{\partial q_i}{\partial q_j^B} =
\frac{\partial q_i}{\partial (q_j + B_j)} = \frac{\partial
q_i}{\partial q_j}\times \left(\frac{\partial (q_j +
B_j)}{\partial q_j}\right)^{-1} = \delta_{ij}.
\label{tranjac}
\end{equation} 
Hence,
$\mathbf{J}^B=I_N$, where $I_N$ is the $N\times N$ identity
matrix. The transformation factor is therefore given by
\begin{equation}
(p^B|p)=\frac{\bm{p}^{\mathsf{T}}I_N^{\mathsf{T}}I_N\bm{p}}
{\bm{p}^{\mathsf{T}}\bm{p}}= 1.
\label{tranfact}
\end{equation}
\subsubsection{Dilatation} The Jacobian for a dilatation is
\begin{equation}
\label{diljac} J^A_{ij}=\frac{\partial q_i}{\partial {q_j^A}} =
\frac{\partial q_i}{\partial (Aq_j)}=A^{-1}\delta_{ij}.
\end{equation}
Then $\mathbf{J}^A=A^{-1}I_N$ and
\begin{equation}
\label{dilfact}
(p^A|p)=\frac{\bm{p}^\mathsf{T}A^{-1}I_N^{\mathsf{T}}A^{-1}I_N
\bm{p}}{\bm{p}^{\mathsf{T}}\bm{p}} = A^{-2},
\end{equation}
or
\begin{equation}
\label{dilfact2} (p|p^A)=A^2.\end{equation}
\subsubsection{Rotation}
We have \begin{eqnarray} \nonumber q^R&=&Rq\\
\nonumber \Rightarrow\qquad q&=&R^{-1}q^R\\
\nonumber \mathrm{and}\qquad
q_i&=&\sum_k\left(R^{-1}\right)_{ik}q^R_k,\\
\nonumber \mathrm{so\;that}\qquad J^R_{ij}=\frac{\partial
q_i}{\partial q^R_j} &=&\left(R^{-1}\right)_{ij}.\end{eqnarray}
Furthermore, since $R$ is orthogonal
\begin{eqnarray}
\nonumber R^{\mathsf{T}}R=RR^{\mathsf{T}}&=&I_N\\
\nonumber \Rightarrow\qquad  R^{\mathsf{T}} &=& R^{-1}\\
\nonumber \mathrm{so\;that}\qquad \mathbf{J}^R =
R^{\mathsf{T}}\qquad&\mathrm{and}&\qquad{\mathbf{J}^R}^{\mathsf{T}}=R.
\end{eqnarray}
This leads to
\begin{equation}
\label{rotfact} (p^R|p)=\frac{\bm{p}^\mathsf{T}RR^{\mathsf{T}}
\bm{p}}{\bm{p}^{\mathsf{T}}\bm{p}}=\frac{\bm{p}^\mathsf{T}I_N
\bm{p}}{\bm{p}^{\mathsf{T}}\bm{p}}=1.
\end{equation}
\subsubsection{Inversion} This requires a modicum of care. First,
observe that
\begin{eqnarray}
\nonumber \frac{\partial r^2}{\partial
q_i}&=&\frac{\partial}{\partial q_i}\sum_k q_k^2\\
\nonumber \Rightarrow \qquad 2r\frac{\partial r}{\partial
q_i}&=&2q_i\\
\label{rdiff} \Rightarrow \qquad \frac{\partial r}{\partial q_i}
&=& \frac{q_i}{r}.
\end{eqnarray}
For the Jacobian we get
\begin{eqnarray}
\nonumber J^{\ast}_{ij}=\frac{\partial q_i}{\partial q^{\ast}_j}
&=&\left(\frac{\partial q^{\ast}_j}{\partial q_i}\right)^{-1}\\
\nonumber \Rightarrow\qquad \left(J^{\ast}_{ij}\right)^{-1}
=\frac{\partial q^{\ast}_j}{\partial
q_i}&=&\frac{\partial}{\partial q_i}\left(\frac{q_j}{r^2}\right)\\
\nonumber &=&
\frac{\delta_{ij}}{r^2}-\frac{2q_j}{r^3}\frac{\partial r}{\partial
q_i}\\
\label{invjac} &=&\frac{\delta_{ij}}{r^2}-\frac{2}{r^4}q_iq_j.
\end{eqnarray}
Now, using (\ref{vswitch})
\begin{equation}
\label{ast}
(p|p^{\ast})=\frac{{\bm{p}^{\ast}}^{\mathsf{T}}
{{\mathbf{J}^{\ast}}^{-1}}^{\mathsf{T}}
{\mathbf{J}^{\ast}}^{-1}\bm{p}^{\ast}}
{{\bm{p}^{\ast}}^{\mathsf{T}}\bm{p}^{\ast}},
\end{equation}
where
\begin{eqnarray}
\nonumber [{{\mathbf{J}^{\ast}}^{-1}}^{\mathsf{T}}
{\mathbf{J}^{\ast}}^{-1}]_{kj}=\sum_i\left[J^{\ast}_{ik}
J^{\ast}_{ij}\right]^{-1}&=&\sum_i\left(\frac{\delta_{ik}}{r^2}
-\frac{2}{r^4}q_iq_k\right)\left(\frac{\delta_{ij}}{r^2}
-\frac{2}{r^4}q_iq_j\right)\\
\nonumber &=&\sum_i\frac{\delta_{ik}\delta_{ij}}{r^4}
-\frac{2}{r^6}\sum_iq_i(q_j\delta_{ik} + q_k\delta_{ij}) +
\frac{4}{r^8}\sum_iq_i^2q_kq_j\\
\nonumber&=&\frac{\delta_{kj}}{r^4} - \frac{2}{r^6}q_jq_k -
\frac{2}{r^6}q_kq_j + \frac{4r^2}{r^8}q_kq_j\\
\nonumber &=& \frac{\delta_{kj}}{r^4}\\
\label{sqre} \Rightarrow \qquad
{{\mathbf{J}^{\ast}}^{-1}}^{\mathsf{T}}{\mathbf{J}^{\ast}}^{-1}
&=&r^{-4}I_N.
\end{eqnarray}
Substituting this into (\ref{ast}) gives us
\begin{equation}
(p|p^{\ast})=\frac{{\bm{p}^{\ast}}^{\mathsf{T}}
r^{-4}I_N\bm{p}^{\ast}}{{\bm{p}^{\ast}}^{\mathsf{T}}\bm{p}^{\ast}}
~=~ r^{-4}~~~~~~~~
\Rightarrow~~~~\qquad (p^{\ast}|p)~=~ r^4.
\label{invfact}
\end{equation}

\section{Inhomogeneous Term}

In this section we prove the invariance of the inhomogeneous term, 
and hence of the cocycle condition under $D$--dimensional 
M\"obius transformations. 

\subsection{Translation} 
The cocycle condition (\ref{cocycle}) can be written
\begin{equation}
\label{useful} (q^a;q^b)=(p^b|p^c)(q^a;q^c) + (q^c;q^b),
\end{equation}
which tells us that
\begin{eqnarray}
\nonumber (q+B+C;q)&=&(p|p^B)(q+B+C;q+B)+(q+B;q)\\
\label{BCb}&=&(q+B+C;q+B) + (q+B;q),
\end{eqnarray}
or, 
\begin{eqnarray}
\nonumber (q+B+C;q)&=&(p|p^C)(q+B+C;q+C)+(q+C;q)\\
\label{BCc}&=&(q+C+B;q+C) + (q+C;q),
\end{eqnarray}
where $B,C$ are two arbitrary constant vectors. Now consider
restricting $B,C$ so that they each have only a single component
along the $j$-axis:
$$B=(0,0,\ldots,0,B_j,0,\ldots,0)\qquad \mathrm{and}\qquad
C=(0,0,\ldots,0,C_j,0,\ldots,0)\; ;\qquad j\in\{1,\ldots ,N\}.$$
We can then define a function
\begin{equation} \label{eff} f(B,q)\equiv (q+B;q),
\end{equation}
and, equating (\ref{BCb}) with (\ref{BCc}) we obtain
\begin{eqnarray}
\nonumber f(C,q+B) + f(B,q) &=& f(B,q+C) + f(C,q)\\
\label{fcb}\Rightarrow\qquad f(C,q+B) &=& f(B,q+C) + f(C,q) -
f(B,q).
\end{eqnarray}
We can differentiate this with respect to $B_j$. Note first that
$$
\frac{\partial (q_j + B_j)}{\partial q_j} = \frac{\partial(q_j +
B_j)}{\partial B_j} = 1,
$$
so for any function $P$ of the combination $q+B$ we have
\begin{eqnarray}
\nonumber
\partial_{B_j}P(q+B)&=&\frac{\partial P(q+B)}
{\partial q_j}\times\left(\frac{\partial(q_j + B_j)}{\partial q_j}
\right)^{-1}\times \frac{\partial (q_j + B_j)}{\partial B_j}\\
\label{Bdiff}&=&\partial_{q_j}P(q+B)\; ; \qquad
\mathrm{where}\qquad
\partial_z\equiv \frac{\partial}{\partial z}.
\end{eqnarray}
Differentiating (\ref{fcb}) then:
\begin{equation}
\label{fdiff}
\partial_{q_j}f(C,q+B)=\partial_{B_j}\left[f(B,q+C)
-f(B,q)\right].
\end{equation}
From (\ref{same0})
\begin{equation}
\label{fvanish}f(0,q)\equiv(q;q) = 0.
\end{equation}
We can
express $f$ in a general series form which guarantees this. Writing 
\begin{equation}
\label{fseries} f(B,q)=\sum_{n=1}^\infty c_n(q)B_j^n.
\end{equation}
and substituting (\ref{fseries}) into (\ref{fdiff}) we have
\begin{eqnarray}
\nonumber
\partial_{q_j}\sum_{n=1}^{\infty}c_n(q+B)C_j^n
&=&\partial_{B_j}\sum_{n=1}^{\infty}\left[c_n(q+C)
-c_n(q)\right]B_j^n\\
\label{seriessub}&=&\sum_{n=1}^\infty\left[c_n(q+C)
-c_n(q)\right]nB_j^{n-1}.
\end{eqnarray}
Using a Taylor expansion the term on the left hand side becomes
\begin{eqnarray}
\nonumber\partial_{q_j}\sum_{n=1}^\infty c_n(q+B)C_j^n
&=&\partial_{q_j} \sum_{n=1}^\infty \left[\sum_{m=0}^\infty
\frac{1}{m!}\partial_{q_j}^m c_n(q)B_j^m\right]C_j^n\\
\nonumber &=& \sum_{n=1}^\infty\left[\sum_{m=0}^\infty
\frac{1}{m!}\partial_{q_j}^{m+1} c_n(q)B_j^m\right]C_j^n\\
\nonumber &=& \sum_{n=1}^\infty\left[\sum_{l=1}^\infty
\frac{1}{(l-1)!}\partial_{q_j}^{l} c_n(q)B_j^{l-1}\right]C_j^n\; ,
\qquad
\mathrm{where}\;\; l=m+1,\\
\label{nswitch} &=&\sum_{l,n=1}^\infty
\frac{1}{(n-1)!}\partial_{q_j}^{n} c_l(q)B_j^{n-1}C_j^l,
\end{eqnarray}
where in the last line we have simply switched indices ($n$ for
$l$) under the double summation. The right hand side of
(\ref{seriessub}) becomes:
\begin{eqnarray}
\nonumber \sum_{n=1}^\infty\left[c_n(q+C)
-c_n(q)\right]nB_j^{n-1}&=&\sum_{n=1}^\infty
\left[\sum_{l=0}^\infty\frac{1}{l!}\partial_{q_j}^lc_n(q)C_j^l
- \frac{1}{0!}\partial_{q_j}^0c_n(q)C_j^0\right]nB_j^{n-1}\\
\label{secondB} &=&
\sum_{l,n=1}^\infty\frac{1}{l!}\partial_{q_j}^lc_n(q)C_j^lnB_j^{n-1}.
\end{eqnarray}
Equating (\ref{nswitch}) with (\ref{secondB}) we find
\begin{eqnarray}
\nonumber\sum_{l,n=1}^\infty \frac{1}{(n-1)!}\partial_{q_j}^{n}
c_l(q)B_j^{n-1}C_j^l &=&
\sum_{l,n=1}^\infty\frac{n}{l!}\partial_{q_j}^lc_n(q)B_j^{n-1}C_j^l\\
\label{crel} \Rightarrow\qquad \frac{1}{n!}\partial_{q_j}^{n}
c_l(q)&=& \frac{1}{l!}\partial_{q_j}^lc_n(q).
\end{eqnarray} Setting $l=1$ gives
\begin{equation}
\label{l1}
\partial_{q_j}c_n(q)=\frac{1}{n!}\partial_{q_j}^nc_1(q).
\end{equation}
If we differentiate (\ref{fseries}) with respect to $q_j$ and
substitute in (\ref{l1}) we obtain
\begin{eqnarray}
\nonumber \partial_{q_j}f(B,q)=\sum_{n=1}^\infty
\partial_{q_j}c_n(q)B_j^n&=&
\sum_{n=1}^\infty\frac{1}{n!}\partial_{q_j}^nc_1(q)B_j^n\\
\nonumber &=& \sum_{n=0}^\infty \frac{1}{n!} \partial_{q_j}^n
c_1(q) B_j^n -
c_1(q)\\
\label{minus} &=& c_1(q+B) -c_1(q).
\end{eqnarray}
Integrating this:
\begin{equation}
\label{int1}
f(B,q)=c(q+B)
-c(q)+g(B,\hat{q}),
\end{equation}
where we define $c(q)$ such that $\partial_{q_j}c(q)=c_1(q)$ and
where $g(B,\hat{q})$ is some function of anything \emph{but} $q_j$
(so $\hat{q}$ denotes all the $q_i$ with $i\neq j$). Recall from
(\ref{fvanish}) that $f(0,q)=0$. Thus:
\begin{equation}
\label{g0van}
 f(0,q)=c(q) - c(q) + g(0,\hat{q}) = g(0,\hat{q}) =
0.
\end{equation}
Using the cocycle condition (\ref{BCb}) once again:
\begin{eqnarray} \label{cocyc}
(q+B+C;q)&=&(q+B+C;q+B) + (q+B;q)\\
\nonumber \Rightarrow\qquad f(B+C,q)&=&f(C,q+B) + f(B,q),
\end{eqnarray}
that is
$$
 c(q+B+C) -c(q) + g(B+C,\hat{q})=c(q+B+C) -c(q+B) +
g(C,\hat{q} + \hat{B}) + c(q+B) - c(q) + g(B,\hat{q}).
$$
 Also
$\hat{B}=0$ since by construction $B_i=0$ for $i\neq j$. So we end
up with\begin{equation} \label{glin2} g(B+C,\hat{q}) =g(C,\hat{q})
+ g(B,\hat{q}).
\end{equation}
We have shown that $g$ is linear in its first argument, and that
it vanishes when its first argument is set to zero. We can write
$g$ in a general form which guarantees these properties:
\begin{equation}
\label{geng} g(B,\hat{q})=K(\hat{q})B_j.
\end{equation}
Now differentiating $f(B,q)$, with respect to $B_j$ this time, we
find on the one hand
\begin{eqnarray}
\nonumber \partial_{B_j}f(B,q)&=&\partial_{B_j}\left[c(q+B) - c(q)
+ g(B,\hat{q})\right]\\
\nonumber &=&\partial_{q_j}c(q+B) + \partial_{B_j}K(\hat{q})B_j\\
\label{gdiff} &=&c_1(q+B) + K(\hat{q}),
\end{eqnarray}
and on the other hand,
\begin{equation}
\label{gdiff2} \partial_{B_j}f(B,q)=\sum_{n=1}^\infty
nc_n(q)B_j^{n-1} = c_1(q) + 2c_2(q)B_j + \ldots,
\end{equation}
so setting $B=0$ and equating (\ref{gdiff}) with (\ref{gdiff2}) we
have
\begin{eqnarray}
\nonumber c_1(q) + K(\hat{q}) &=& c_1(q)\\
\label{K0} \Rightarrow\qquad K(\hat{q})&=&0.
\end{eqnarray}
This means that
$$
g(B,\hat{q}) = 0\times B_j = 0,
$$
and so we arrive at the result
\begin{equation}
\label{impf} f(B,q)=c(q+B)-c(q).
\end{equation}
Next, we consider a general constant vector $D$ (which we need not
restrict to having only a single component). We first define the
function
\begin{equation}\label{Gdef} G(D,q)\equiv(q+D;q).\end{equation}
Note that we must have
\begin{equation}\label{Greduce}G(B,x)=f(B,x)\end{equation} where as before $B$
has only one component along the $j$-axis. Again we use the
cocycle condition (\ref{BCb},\ref{BCc}),
\begin{eqnarray}
\nonumber(q+B+D;q)&=&(q+B+D;q+B)+(q+B;q)\\
\nonumber\emph{and}\qquad (q+B+D;q)&=&(q+D+B;q+D) + (q+D;q)\\
\nonumber\Rightarrow (q+B+D;q+B)-(q+B+D;q+D)&=&(q+D;q) -
(q+B;q)\\
\nonumber \mathrm{so}\qquad G(D,q+B) -G(B,q+D) &=& G(D,q) - G(B,q),\\
\nonumber \mathrm{that\;is}\qquad G(D,q+B) - f(B,q+D) &=& G(D,q) -
f(B,q)\\
\nonumber \Rightarrow G(D,q+B)-c(q+D+B)+c(q+D)&=&G(D,q)-c(q+B) +
c(q).
\end{eqnarray}
Differentiating this with respect to $B_j$ we get
\begin{equation}
\label{diffD}
\partial_{q_j}G(D,q+B) - \partial_{q_j}c(q+D+B) =
-\partial_{q_j}c(q+B),
\end{equation}
which yields, for $B=0$,
\begin{equation}
\label{diffD2}
\partial_{q_j}G(D,q)=\partial_{q_j}\left[c(q+D) - c(q)\right].
\end{equation}
Integrating \emph{this} gives
\begin{equation}
\label{Ghat}G(D,q)=c(q+D)-c(q) + \hat{G}(D,\hat{q}),
\end{equation}
where $\hat{G}(D,\hat{q})$ is some function of the $q_i$ ($i\neq
j$) which by (\ref{Greduce}) and (\ref{impf}) we know must vanish
if $D$ has only a single component along the $j$-axis. Putting
this back into the cocycle condition again we get
\begin{eqnarray}
\label{Gco} G(D+B,q)&=&G(B,q+D) + G(D,q)\\
\nonumber \Rightarrow c(q+D+B) - c(q)
+\hat{G}(D+B,\hat{q})&=&c(q+D+B)-c(q+D) +
\hat{G}(B,\hat{q}+\hat{D}) +\\
\nonumber && c(q+D)-c(q)+\hat{G}(D,\hat{q})\\
\nonumber \mathrm{so}\qquad \hat{G}(D+B,\hat{q}) &=&
\hat{G}(B,\hat{q}+\hat{D}) + \hat{G}(D,\hat{q}),
\end{eqnarray}
so we see that $\hat{G}(D,\hat{q})$ satisfies the same algebra as
$G(D,q)$ (only with one fewer variable since $\hat{q}$ excludes
$q_j$). We can therefore apply the same arguments to
$\hat{G}(D,\hat{q})$ as we have used for $G(D,q)$: We'll end up
with a relation like (\ref{Ghat}) for $\hat{G}(D,\hat{q})$, with a
new function tacked onto the end which is now only a function of
$N-2$ of the $q_i$. Then we put \emph{this} back into the cocycle
condition - and so on and so on...

 Having worked our way
recursively through all $N$ components we obtain
\begin{equation}
\label{Gprop} G(D,q)\equiv(q+D;q)=F(q+D)-F(q) + H(D),
\end{equation}
where $H(D)$ is some function which vanishes whenever $D$ has only
one component. Since $H(D)$ has no second argument, we find that
upon substitution of (\ref{Gprop}) into (\ref{Gco}) we are left
with
\begin{equation}
\label{Hlin} H(D+C) = H(D) + H(C), \end{equation} so $H$ is
linear. We can write $H$ in a general form which guarantees this
property:
\begin{equation}
\label{Hlincomb} H(D)=\sum_{i=1}^Na_iD_i
\end{equation}
(just a linear combination of the components of $D$). However we
require that $H=0$ if $D_i=0$ and $D_j\neq 0$ ($i\neq j$) for any
$j$, so we must have that $a_j=0$ for any $j$, which means that
$H$ is identically vanishing. We have determined the
form for the inhomogeneous term under arbitrary translations to be
\begin{equation}
(q+D;q)=F(q+D)-F(q).
\label{finally} 
\end{equation}
\subsection{Dilatation} In a similar way to our previous treatment
we define a function
\begin{equation}
\label{hdef} h(A,q)\equiv(Aq;q),
\end{equation}
and use the cocycle condition to examine its structure. From
(\ref{useful}), and using (\ref{finally}), we have
\begin{eqnarray}
\nonumber(A[q+B];q)&=&(p|p^A)(A[q+B];Aq)+(Aq;q)\\
\nonumber &=&A^2(Aq + AB;Aq) + (Aq;q)\\
\label{A1} &=&A^2\left[F(Aq +AB)-F(Aq)\right] + h(A,q), 
\end{eqnarray}
which may also be written as, 
\begin{eqnarray}
\nonumber(A[q+B];q)&=&(p|p^B)(A[q+B];q+B) + (q+B;q)\\
\label{A2} &=&h(A,q+B) + F(q+B) - F(q).
\end{eqnarray}
Equating Eqs. (\ref{A1}) and \ref{A2}) we obtain,
\begin{equation}
\label{heq}
h(A,q+B)-h(A,q)=A^2\left[F(Aq+AB)-F(Aq)\right]-F(q+B) +
F(q),
\end{equation}
and differentiating (\ref{heq}) with respect to $B_j$ we find
\begin{equation}
\label{hdiff1}
\partial_{q_j}h(A,q+B)=A^2\partial_{q_j}F(Aq+AB) -
\partial_{q_j}F(q+B).
\end{equation}
Setting $B=0$ gives
\begin{equation}
\label{B0hdiff}
\partial_{q_j}h(A,q)=\partial_{q_j}\left[A^2F(Aq)-F(q)\right],
\end{equation}
which upon integration yields
\begin{equation}
\label{inth} h(A,q)=A^2F(Aq) - F(q) + g(A),
\end{equation}
where $g(A)$ is some function which cannot depend on any of the
$q_i$ (as we could have chosen any $j$ in (\ref{hdiff1})). Now,
at the origin ($q=0$) a dilatation can have no effect, 
so $h(A,0)$ is independent of $A$ and
\begin{eqnarray}
\nonumber h(A,0)=h(1,0)\equiv(q;q)&=&0\\
\nonumber \Rightarrow h(A,0) = A^2F(A\times 0) - F(0) +
g(A) &=& 0\\
\nonumber \Rightarrow \qquad g(A)=F(0)\left[1-A^2\right].
\end{eqnarray}
Putting this back into (\ref{inth}) gives
\begin{equation}
h(A,q)=A^2\left[F(Aq) - F(0)\right] - \left[F(q) - F(0)\right],
\end{equation}
or more concisely
\begin{equation}
\label{Dilresult} h(A,q)\equiv(Aq;q)=A^2F(Aq)-F(q),
\end{equation}
where we defined $F(0)=0$.
\subsection{Rotation} 
Following similar steps, from the cocycle condition
(\ref{A1},\ref{A2}) we have
\begin{eqnarray}
\label{R1} (R[q+B];q)&=&(p|p^R)(Rq+RB];Rq) + (Rq;q)\\
\label{R2}\mathrm{and}\qquad
(R[q+B];q)&=&(p|p^B)(R[q+B];q+B)+(q+B;q)\\
\label{rotso}\Rightarrow (R[q+B];q+B)-(Rq;q)&=&(Rq + RB;Rq)
-(q+B;q).\\
\nonumber \mathrm{Using\;(\ref{finally}),}\qquad (R[q+B];q+B) -
(Rq;q)&=&\left\{F(R[q+B])-F(q+B)\right\}-\left\{F(Rq)-F(q)\right\}.
\end{eqnarray}
This is satisfied by
\begin{equation}
\label{Rcon} (Rq;q)=F(Rq) - F(q) + C\; ; \qquad
C\;\mathrm{a\;constant}.
\end{equation}
A rotation can have no effect at the
origin, so $(Rq;q)|_{q=0}\equiv
(q;q)=0$ and
\begin{equation}
\nonumber F(R\times 0)-F(0) + C = 0~~~~\Rightarrow~~~~ \qquad C=0.
\end{equation}
The form of the inhomogeneous term under rotations is therefore
$(Rq;q)=F(Rq)-F(q).$

\subsection{Inversion} To begin with we consider how lengths
transform under the generators of the M\"{o}bius group.
\begin{equation}
\label{arestar} {r^{\ast}}^2=\sum_{i=1}^N(q^{\ast})_i^2=
\sum_{i=1}^N\left(\frac{q_i}{r^2}\right)^2=\frac{1}{r^4}r^2=\frac{1}{r^2},
\end{equation}
\begin{equation}
\label{rA} {r^A}^2=\sum_{i=1}^N(Aq)_i^2=A^2r^2, \end{equation} and
trivially, since lengths are preserved by rotations,
\begin{equation}
\label{rR} {r^R}^2=r^2. \end{equation}
 (\ref{arestar}) fixes $q$ as
\emph{involutive}:
\begin{equation}
\label{invol} (q^{\ast})_i^{\ast}=
\left(\frac{q}{r^2}\right)_i^{\ast}
=\frac{q_i^{\ast}}{{r^{\ast}}^2}
=\frac{q_i}{r^2\frac{1}{r^2}}=q_i.
\end{equation}
From (\ref{rR}) we see that rotation commutes with inversion:
\begin{equation}
\label{rotinv} (Rq)_i^{\ast}=\frac{(Rq)_i}{{r^R}^2}=\frac{\sum_k
R_{ik}q_k}{r^2}=\sum_k R_{ik}q_k^{\ast}=(Rq^{\ast})_i.
\end{equation}
Finally, under dilatations we have, using (\ref{rA}),
\begin{equation}
\label{dilinv} (Aq)_i^{\ast}=\frac{(Aq)_i}{{r^A}^2}=
\frac{Aq_i}{A^2r^2}=A^{-1}q_i^{\ast}.
\end{equation}
Now, applying (\ref{swap}), we see from (\ref{invol}) that
\begin{equation}
\label{invvan}
(q^{\ast};q)=-(p|p^{\ast})(q;q^{\ast})=-\frac{1}{r^4}([q^{\ast}]^{\ast};q^{\ast}),
\end{equation}
which vanishes when evaluated at any $q$ such that $q=q^{\ast}$:
\begin{equation}
\label{thatis} (q^{\ast};q)|_{q=q_0}=0\; ;\qquad q_0=q_0^{\ast}.
\end{equation}
Next, bearing in mind (\ref{Dilresult}), we revisit the cocycle
condition (\ref{useful}):
\begin{eqnarray}
\nonumber ([Aq]^{\ast};q)&=&(p|p^A)([Aq]^{\ast};Aq)+(Aq;q)\\
\label{invfirst} &=&A^2([Aq]^{\ast};Aq) + A^2F(Aq)-F(q),
\end{eqnarray}
but using (\ref{dilinv}), we can equally write
\begin{eqnarray}
([Aq]^{\ast};q)&=&(p|p^{\ast})([Aq]^{\ast};q^{\ast}) +
(q^{\ast};q)\\
\nonumber &=& \frac{1}{r^4}(A^{-1}q^{\ast};q^{\ast}) +
(q^{\ast};q)\\
\label{invsec} &=&
\frac{1}{r^4}\left[(A^{-1})^2F(A^{-1}q^{\ast})-F(q^{\ast})\right]
+ (q^{\ast};q).
\end{eqnarray}
So, equating (\ref{invfirst}) with (\ref{invsec}) gives:
\begin{equation}
\label{invec} A^2([Aq]^{\ast};Aq) +
A^2F(Aq)-F(q)=\frac{1}{r^4}\left[A^{-2}F(A^{-1}q^{\ast})-F(q^{\ast})\right]
+ (q^{\ast};q).
\end{equation}
If we choose a point $q_0$ on the surface of the unit sphere, so
that $r_0^2=\sum_i({q_0}_i)^2=1$ and $q_0=q_0^{\ast}$, then (by
(\ref{thatis})) (\ref{invec}) reduces to
\begin{eqnarray}
\nonumber A^2([Aq_0]^{\ast};Aq_0)&=&A^{-2}F(A^{-1}q_0^\ast) -
F(q_0^\ast) -
A^2F(Aq_0) + F(q^0)\\
\label{near}\Rightarrow\qquad
([Aq_0]^{\ast};Aq_0)&=&\frac{1}{A^4}F(A^{-1}q_0^\ast)-F(Aq_0).
\end{eqnarray}
Lastly, we note that any vector $q$ can be expressed in the polar
form $r\hat{q}$ where $r$ is the length of $q$ and where
$\hat{q}$ is a vector of unit length parallel to $q$. 
Clearly $\hat{q}$ automatically
has the property required of $q_0$. The mapping
$\hat{q}\rightarrow r\hat{q}$ is simply a dilatation of
$\hat{q}$ with $A=r$, so we can put $Aq_0=q$ (and from
(\ref{dilinv}) $A^{-1}q_0^{\ast}=[Aq_0]^{\ast}=q^{\ast}$) in
(\ref{near}) to obtain the final result,
\begin{equation}
\label{Invres} (q^{\ast};q)=\frac{1}{r^4}F(q^{\ast})-F(q).
\end{equation}
\section{Summary}
To summarise the quantum Hamilton--Jacobi, eq. (\ref{QSHJE}),
and the cocycle condition, eq. (\ref{cocycle}), 
imply that $(q^a; q^b)$ vanishes if $q^a$ and $q^b$ are 
related by a M\"obius transformation, that is 
\beqn
(q+B;q) & = &0, \label{eqq18}\\ 
(Aq;q)  & = &0, \label{eqq17}\\ 
(\Lambda q;q) & = & 0, \label{rotazioni}\\  
(q^*;q)       & = & 0. \label{inversione2}
\eeqn

These equations are equivalent to $(\gamma(q);q)=0$, where $\gamma(q)$ is a 
general M\"obius transformation. Moreover, from eq. (\ref{cocycle}) we have 
\beq
(\gamma(q^a);q^b)=(q^a;q^b),\qquad 
(q^a;\gamma(q^b))=(p^{\gamma(b)}|p^b) (q^a;q^b). 
\label{cs}
\eeq
Considering the Jacobian factor $(p^{\gamma(b)}|p^b)$ we0 observe that the 
M\"obius transformation is conformal with respect to the Euclidean metric, 
{\it i.e.}00
\beq
ds^2=\sum_{j=1}^D d\gamma(q)_jd\gamma(q)_j=\sum_{j,k,l=1}^D{\partial\gamma(q)_j 
\over\partial q_k}{\partial\gamma(q)_j\over\partial q_l}dq_kdq_l=\sum_{j=1}^D 
e^{\phi_\gamma(q)}dq_jdq_j. 
\label{conf1}
\eeq 
Hence,
\beq
(p^{\gamma(b)}|p^b)=e^{-\phi_{\gamma}(q^b)}. 
\label{conf2}
\eeq 
We note that in the case of rotations and translations the conformal 
re--scaling is the identity. 
For dilatations $\exp\phi^A=A^2$, whereas for the inversion 
$\exp\phi^*=r^{-4}$. 
 
We remark that this conformal structure is obtained by setting 
$S_0^v(q^v)=S_0(q)$. We would like to emphasise that this is not 
a restriction on the formalism, but merely a convenient choice. 
Any transformation that we may impose, other than 
$S_0^v(q^v)=S_0(q)$, would yield the same results. 
The freedom in setting $S_0^v(q^v)=S_0(q)$ 
results from the fact that $q$ and $q^v$ 
represent the spatial coordinates in their own systems. 
The condition $S_0^v(q^v)=S_0(q)$ can be 
seen just as the simplest way to set the coordinate transformations from 
the system with 
reduced action $S^v_0$ to the one with reduced action 
$S_0$. Since the physics is determined by the functional 
structure of 
$S^v_0$, we can denote the coordinate as we like. 
However, this is not the case in classical mechanics, as 
for a free particle with vanishing energy we have 
$S_0(q)=cnst$, and imposing $S_0^v(q^v)=S_0(q)$
does not make sense. 
Requiring that this is well--defined for any system is 
synonymous to imposing the equivalence postulate, 
and the definability of phase space duality for all physical states. 
The existence of the conformal structure, manifested by the invariance
of the inhomogeneous term under M\"obius transformations, 
which in  $D \geq3$ coincides with the conformal group, 
is at the core of quantum mechanics.

\section{Quadratic identities} 

In the previous sections we discussed the cocycle condition, 
which is obtained by requiring the QHJE equation retain
its form under coordinate transformations. The transformation
properties of the kinetic term fix those of the classical 
and added potential to transform as quadratic differentials. 
Furthermore, in the one dimensional case the M\"obius 
symmetry uniquely fixes the functional form of the 
inhomogeneous term to that of the Schwarzian derivative. 
An identity of Schwarzian derivatives follows from 
these transformation properties and is given by
$${\left({{\partial S_0}%
\over{\partial q}}\right)^2={\beta^2\over{2}}\left(\{{\rm e}^{{i2S_0}\over%
\beta};q\}-\{S_0;q\}\right)}$$
With the identifications 
\begin{eqnarray}
W(q) & = & -{\beta^2\over{4m}}\{{\rm e}^{{i2S_0}\over\beta};q\}=V(q)-E\label{wqid}\\
Q(q) & = & {\beta^2\over{4m}}\{S_0;q\},\label{qqid}
\end{eqnarray}
the modified Hamilton--Jacobi equations becomes 
$${1\over{2m}}\left({{\partial S_0}\over{\partial q}}\right)^2+%
V(q)-E+{\beta^2\over{4m}}\{S_0;q\}=0.$$
We therefore note that for the state $W(q)\equiv 0$
the Modified HJ equation admits the solutions 
$${S}_0~  =  ~\pm{\beta\over2}\ln {q^0}~
\ne~A { q^0} + B. $$
Hence, quantum mechanics enables consistency of the equivalence postulate 
by removing the linear solutions from the space of solutions. As we discussed
before, since $Q(q)$ is never vanishing quantum mechanics carries within
it its own regularisation scheme. Furthermore, from eq. (\ref{wqid})
and the properties of the Schwarzian derivative,
the function $W(q)=V(q)-E$ is a potential of a second order
differential equation given by, 
$$\left(-{\beta^2\over{2m}}{\partial^2\over{\partial q^2}}+V(q)-E\right)%
\Psi(q)=0,$$
which is the Schr\"odinger equation and we may 
identify the covariantising parameter $\beta$ with 
the Planck constant $\hbar$. The general solutions is given by
$$
\Psi(q)= \left(A\psi_1+B\psi_2\right)=
{1\over{\sqrt{S_0^\prime}}}\left(A{\rm e}^{+{i\over\hbar}S_0}+
B{\rm e}^{-{i\over\hbar}S_0}\right).$$
The solution of the Schwarzian equation eq. (\ref{wqid}) is then 
given in terms of ratio of the solutions of the 
Schr\"odinger equation, {\it i.e.}
$${\rm e}^{+{{i2S_0}\over\hbar}}~=~{\rm e}^{i\alpha}~
{{w+i{\bar\ell}}\over{w-i\ell}}~~~~~~\hbox{where}~~~~~~~~w~=~{\psi_1\over\psi_2},$$
where, due to the symmetries of the Schwarzian derivative,
the solution is given up to a M\"obius transformation. 
Furthermore, from the condition that 
$S_0(q)\ne constant$ we have that the constants $\ell$ and $\alpha$
satisfy $\ell=\ell_1+i\ell_2$, $\ell_1\ne0$ and $\alpha\in R$.

While appearance of the Schwarzian derivative in the one dimensional case
may seem a bit esoteric, the multi--dimensional case reveals more
clearly the simplicity of the formalism. Consider applying the 
Laplacian to the function 
$$\psi=R(q){\rm e}^{\alpha S_0(q)}, ~~~~{\it i.e.}~~~~
\Delta \left(R(q) {\rm e}^{\alpha S_0(q)}\right)\, .$$
Proper application of the chain rule then leads to a quadratic 
identity given by
\beq
\alpha^2(\nabla S_0)\cdot(\nabla S_0)=%
{\Delta(Re^{\alpha S_0})\over Re^{\alpha S_0}}%
-{\Delta R\over R}-
{\alpha\over R^2}\nabla\cdot\left({R^2\nabla S_0}\right).
\label{did}
\eeq
Setting $\alpha=i/\hbar$ the imaginary part of eq. (\ref{did})
gives a continuity equation. The first term on the right--hand side
of eq. (\ref{did}) is identified with the classical potential,
yielding the $D$--dimensional nonrelativistic Schr\"odinger 
equation, with the general solution given by
\beq
\Psi(q)= \left(A\psi_1+B\psi_2\right)=
R(q) \left(A{\rm e}^{+{i\over\hbar}S_0}+
B{\rm e}^{-{i\over\hbar}S_0}\right),
\label{bipolar}
\eeq
where $q_i$ are now the $D$--dimensional coordinates. 
These identifications produce the $D$--dimensional stationary
nonrelativistic Quantum Hamilton--Jacobi Equation 
given by
$$
{1\over {2m}}(\nabla S_0)\cdot(\nabla S_0) +V(q)- E -{\hbar^2\over {2m}}
{\Delta R\over R} =0. 
$$

\section{Time parameterisation}

We note that the QHJE is reminiscent of Bohm's approach to quantum 
mechanics \cite{holland}. However, there is a crucial difference,
which is precisely related to the M\"obius symmetry
underlying quantum mechanics in the equivalence postulate
approach. As is well known Bohm's approach argues for the 
existence of a trajectory representation of quantum
mechanics, in which time parameterisation of trajectories
is obtained. In Bohm's approach the wave function is identified 
with $A=0$ and $B\ne 0$ in eq. (\ref{bipolar}). In that case one 
identifies the conjugate momentum as
\beq
p=\hbar {\rm Im}{{\nabla\psi}\over\psi},
\label{polarrep}
\eeq
which can be used to define a trajectory representation 
by identifying the conjugate momentum the mechanical momentum, 
{\it i.e.} by setting $p=m{\dot q}$. However, the choice
$A=0$ and $B\ne 0$ is not consistent with the M\"obius 
symmetry underlying quantum mechanics, which necessitates that
$A\ne $ {\bf and} $B\ne 0$. Alternatively, the M\"obius 
symmetry implies that space is compact, in which case the boundary 
conditions are not compatible with the choice $A=0$ and $B\ne 0$
but impose that both must be included in the solution. 
Therefore, the M\"obius symmetry of the QHJE 
implies that 
$$\nabla S\ne {\hbar}{\rm Im}{{\nabla\psi}\over\psi}$$
and Bohm's definition of trajectories 
is not valid \cite{fmtpqt}. 
An alternative proposal \cite{floyd}
to define trajectories in quantum mechanics
suggests to use Jacobi's theorem that identifies time as the 
derivative of the $S_0$ with respect to $E$, 
and by replacing the solution of the CHJE, with the solution 
of the QHJE, {\it i.e.}
\beq 
t-t_0= {{\partial S_0^{\rm qm}}\over{\partial E}}.  
\label{floydproposal}
\eeq 
Time parameterisation of the trajectory can then be obtained
by inverting $t(q)\rightarrow q(t)$. However, the M\"obius 
symmetry that underlies the QHJE dictates that the energy levels 
are always quantised \cite{fmtpqt}. Hence, differentiation
with respect to the energy is not well defined. Time 
in quantum mechanics can be thought of as a classical 
background parameter, but not as a fundamental quantum
variable. At the quantum level trajectories may only 
have a probabilistic interpretation rather than 
a deterministic representation. It should be stressed, however,
that while a deterministic time parameterisation is not 
consistent with the M\"obius symmetry that underlies the 
QHJE, time parameterisation {\` a} la Bohm \cite{holland}
or via the bi--polar representation \cite{wyatt, poirier}
provides a useful semi--classical approximation.

The compactness of space, imposed by the M\"obius symmetry 
underlying the QHJE, implies that the energy levels are 
always quantised. However, in quantum mechanics this does not 
suffice. The probability interpretation of the wave function 
implies that the wave function for bound states is square integrable. 
In general, the differential equations associated 
with the quantum mechanical problems for bound states 
admit solutions that are not square integrable. In the 
one dimensional case it was proven rigoursly that 
trivialising transformations $q\rightarrow q^0=\psi_1/psi_2$, 
where $\psi_1$ and $\psi_2$ are the two solutions of the 
Schr\"odinger equation, has to be continuous on the 
extended real line, {\it i.e.} the real line
plus the point at infinity. This requirement is synonymous 
to the requirement that the M\"obius symmetry is preserved. 
It is then shown \cite{fmtqt,fmijmp} that this
condition is satisfied iff the corresponding 
Schr\"odinger equation admits a square integrable 
solution. Thus, the same physical states that are 
selected in conventional quantum mechanics by the 
axiomatic probability interpretation of the 
wave function, are selected by consistency
in the EP approach. The EP approach may therefore
be regarded as reproducing the basic properties 
of conventional quantum mechanics, with the caveat that 
spatial space is compact.



\section{Conclusions}

The requirement that the HJ equation retain its form under 
coordinate transformation led to the cocycle condition and 
the QHJE. In turn this led to the removal of the 
classical solution $S_0(q) = const$ from the space 
of admissible solutions and consequently the requirement 
that $p\ne 0$. These properties are intimately related to 
a duality in phase space that is defined in terms of 
the involutive nature of the Legendre transformations \cite{fm2,fmijmp}. 
Now, the Legendre transformations are not defined for 
linear functions. The solutions admitted by the 
QHJE in the case of the trivial classical potential
correspond to the self--dual states under the phase space
duality. Thus, the QHJE enables the consistency 
of the phase space duality for the 
entire space of solutions.  

The M\"obius symmetry underlying the QHJE is the fundamental 
property of quantum mechanics in the equivalence postulate approach.
It provides an alternative to the axiomatic formulation of 
quantum mechanics. It necessitates the existence of 
the quantum potential, which is never vanishing.
The quantum potential may be interpreted as an
internal curvature term of elementary particles 
\cite{fmqt,de}, which can therefore be seen to be 
a direct consequence of the M\"obius symmetry 
underlying the QHJE. In turn, the M\"obius symmetry,
and the duality structure that it enforces,  implies 
the existence of a fundamental length scale 
in the formalism \cite{fmpl, fmijmp}. compatibility
with the classical limit then implies that the 
fundamental length scale may be identified 
with the Planck length. In turn,
the existence of a fundamental length scale 
implies the admission of an ultraviolet cutoff. 
Similarly, the M\"obius symmetry dictates that 
spatial space is compact. Thus, the M\"obius 
symmetry provides an intrinsic quantum mechanical 
regularisation scheme in the ultraviolet, 
as well as the infrared. Furthermore, 
it leads to the phenomenological characteristics 
of quantum mechanics \cite{fmtqt, fmijmp}
without assuming any prior interpretation of the wave function. 
The HJ formalism, augmented with the M\"obius symmetry,
therefore provides an alternative starting point to the 
axiomatic formulation of of quantum mechanics, based 
on the probability interpretation of the wave function. 
In that respect, while a fundamental appreciation of the 
geometrical role of the wave function is yet to be developed,
a key guide may lie in duality relations between 
the wave function and the space coordinates \cite{fm1}.
We further note that the universality of the quantum potential 
implies that it corresponds to a universal force acting
on elementary particles \cite{marco}. 
The M\"obius symmetry underlying quantum 
mechanics implies that spatial space is compact, 
which may have left a remnant in the Cosmic Microwave 
Background Radiation \cite{cmb}. 
Additionally, it leads to modified energy dispersion relations \cite{opera},
which may affect the propagation of gamma rays over 
astrophysical distances \cite{subir}. 

\medskip
{\bf Acknowledgements}

This work is supported in part by the STFC under contract ST/L000431/1.

\end{document}